\documentclass{ws-ijmpe}
\usepackage[super,compress]{cite}

\usepackage{tensor}

\begin{document}

\markboth{S.-I. Ando}{Hypernuclei in Halo/Cluster Effective Field Theory}

\catchline{}{}{}{}{}

\title{
Hypernuclei in Halo/Cluster Effective Field Theory 
}

\author{Shung-Ichi Ando}

\address{
Department of Information Communication \& Display Engineering, \\
Sunmoon University, Asan, Chungnam 336-708, 
Republic of Korea  \\
sando@sunmoon.ac.kr
}

\maketitle

\begin{history}
\received{Day Month Year}
\revised{Day Month Year}
\end{history}

\begin{abstract}
The light double $\Lambda$ hypernuclei, 
${}_{\Lambda\Lambda}^{\ \ 4}$H and 
${}_{\Lambda\Lambda}^{\ \ 6}$He, 
are studied  
as three-body $\Lambda\Lambda d$ and $\Lambda\Lambda\alpha$ cluster systems 
in halo/cluster effective field theory at leading order. 
We find that the $\Lambda\Lambda d$ system in spin-0 channel
does not exhibit a limit cycle whereas the $\Lambda\Lambda d$ system 
in spin-1 channel and the 
$\Lambda\Lambda\alpha$ system in spin-0 channel do.
The limit cycle is associated with the formation of bound states, 
known as Efimov states, 
in the unitary limit. 
For the $\Lambda\Lambda d$ system in the spin-0 channel
we estimate the scattering length $a_0$ for $S$-wave $\Lambda$ 
hyperon-hypertriton scattering as $a_0=16.0\pm 3.0$~fm. 
We also discuss that 
studying the cutoff dependences 
in the $\Lambda\Lambda d$ and $\Lambda\Lambda\alpha$ systems, 
the bound state of ${}_{\Lambda\Lambda}^{\ \ 4}$H is not an Efimov state 
but formed due to a high energy mechanism 
whereas that of ${}_{\Lambda\Lambda}^{\ \ 6}$He may be 
regarded as an Efimov state.

\end{abstract}


\ccode{PACS numbers: 
21.45.-v, 
20.80.+a, 
11.10.Hi}


\section{Introduction}

Light double $\Lambda$ hypernuclei are exotic few-body 
systems that provide an opportunity to investigate 
the $SU(3)$ flavor baryon-baryon interaction in the strangeness $-2$ 
channel~\cite{1,2,3}.
They are also expected to play a key role in resolving the long-standing
puzzle of the existence of the $H$ dibaryon~\cite{jaffe}, 
which has recently been studied 
by lattice QCD simulations~\cite{nplqcd,halqcd}.
Meanwhile $SU(3)$ flavor baryon-baryon potentials
have been obtained in chiral effective theory, 
hyperon-nucleon potentials up to
next-to-leading order~\cite{hpkmnw-npa13} 
and hyperon-hyperon potentials at leading order (LO)~\cite{2}
assuming $SU(3)$ flavor symmetry except for the masses of mesons and 
baryons. The main difficulty of the study is that one does not have 
an adequate number of experimental data to accurately determine
all parameters in the potentials. 
In this situation, it may be worth studying the exotic few-body 
systems by constructing a simple theory 
at very low energies 
which has a handful of parameters.

The first observation of $\nuclide[6][\Lambda\Lambda]{He}$ was reported 
in the 1960s~\cite{Prowse66}, however there have been only a few reports on 
this light hypernucleus~\cite{DDFM89,TAAA01}.
Among them, a track of $\nuclide[6][\Lambda\Lambda]{He}$ was clearly 
caught in an emulsion experiment of the KEK-E373 Collaboration~\cite{TAAA01},
now known as the ``NAGARA" event, and the two-$\Lambda$ separation energy
$B_{\Lambda\Lambda}$ of $\nuclide[6][\Lambda\Lambda]{He}$ is estimated as
$B_{\Lambda\Lambda}=6.93\pm 0.16$~MeV after being averaged with that from the
``MIKAGE'' event~\cite{Nakazawa10,E373-13}.
In addition, the formation of another light double-$\Lambda$ hypernucleus,
${}_{\Lambda\Lambda}^{\ \ 4}$H, is conjectured in the BNL-AGS E906
experiment~\cite{AAAB01}.

Theoretical studies for the double $\Lambda$ hypernuclei 
mainly aim at extracting information on baryon-baryon interactions 
in the strangeness sector and searching for new exotic systems 
for which the value of $B_{\Lambda\Lambda}^{}$
of $\nuclide[6][\Lambda\Lambda]{He}$ plays 
an important role~\cite{HMRY10,Gal10}.
In fact, studies of $\nuclide[6][\Lambda\Lambda]{He}$ 
have been reported with various
issues in theory~\cite{th-prl65,AB67,bim-ptp82,cag-npa97,FG02b},
primarily employing the three-body ($\Lambda\Lambda\alpha$) cluster model.
One of those issues is the importance of a mixing of the $\Xi N$ channel
in the $\Lambda\Lambda$ interaction because the relatively small mass
difference between the $\Xi N$ and $\Lambda\Lambda$ channels is
about 23~MeV (see, e.g., Ref.~\refcite{cag-npa97}).
In theoretical studies for $\nuclide[4][\Lambda\Lambda]{H}$, 
although the first Faddeev-Yakubovsky calculation showed a
negative result~\cite{FG02}, subsequent theoretical
studies~\cite{NAM02,Shoeb04,NSAM04,SUB11}
predicted the possibility of the ${}_{\Lambda\Lambda}^{\ \ 4}$H bound state
based on the phenomenological $\Lambda\Lambda$ potentials which can describe
the bound state of ${}_{\Lambda\Lambda}^{\ \ 6}$He.
In this article, we review the results of our recent 
works on the light double $\Lambda$ hypernuclei, 
$\nuclide[4][\Lambda\Lambda]{H}$ and 
$\nuclide[6][\Lambda\Lambda]{He}$, as 
three-body $\Lambda\Lambda d$ and $\Lambda\Lambda\alpha$ cluster systems,  
studied in halo/cluster effective field 
theories (EFTs) at LO~\cite{dLamH4,dLamHe6}.

EFTs at very low energies are expected to provide a
model-independent and systematic perturbative method where one introduces 
a high momentum separation scale $\Lambda_H$ between relevant degrees 
of freedom in low energy and irrelevant degrees of freedom in high energy 
for the system in question.
Then one constructs an effective Lagrangian expanded in terms of the number of
derivatives order by order.
Coupling constants appearing in the effective Lagrangian should be determined
from available experimental or empirical data.
In the study of the three-body systems,
we will make use of the cyclic singularities 
that arise in the solutions for the 
integral equations in the asymptotic limit~\cite{Danilov61}.
Such singularities are renormalized by introducing a suitably large 
momentum cutoff $\Lambda_c$ ($\Lambda_c \gtrsim \Lambda_H$) 
in the loop integrations at the cost of introducing three-body
counter terms at LO.
Consequently, in order to absorb this cutoff dependence, 
the corresponding three-body coupling may exhibit a cyclic 
renormalization group (RG) evolution termed as the limit cycle~\cite{Wilson71}.
The cyclic singularities are associated with the occurrence of bound states, 
known as the Efimov states, in the resonant/unitary limit~\cite{Efimov71}.
For a review, see, e.g., Refs.~\refcite{BK02,BH04} and references therein.

In our study for the three-body systems, 
we first derive simple equations from the homogeneous 
part of the integral equations in the asymptotic limit 
to make an examination whether
the systems exhibit a limit cycle or not. 
When a limit cycle does not appear in the system, 
it is not necessary to introduce 
a three-body contact interaction 
for renormalization at LO.
The system would be insensitive to the three-body contact term
and can be described by two-body interactions.
In the spin-0 channel of the $\Lambda\Lambda d$ system, 
we shall see that a limit cycle does not appear and thus we can
make a prediction on the scattering length $a_0$ 
for $S$-wave $\Lambda$ hyperon and hypertriton scattering 
using a parameter of a two-body interaction.
When a limit cycle appears in the system, on the other hand,
one needs to introduce a three-body contact interaction at LO. 
Because the cyclic singularity is related to 
the occurrence of Efimov states,
the bound state emerging 
in the system can be regarded as an Efimov state. 
In the study we vary the magnitude of the cutoff within a reasonable 
range, we investigate its effects on the formation of bound states
and discuss whether the bound states of
$\nuclide[4][\Lambda\Lambda]{H}$ and $\nuclide[6][\Lambda\Lambda]{He}$ 
can be Efimov states or not.\footnote{
We refer to an ``Efimov state'' as a bound state forming due to 
a singular interaction which exhibits a limit cycle.
While each of the systems described in the theory has a scale beyond of 
which the applicability of the theory breaks down.
We examine likelihood of the formation of an Efimov state 
in the systems by comparing a cutoff value at which an 
Efimov state emerges with the scale of the theory.
}

This work is organized as follows.
In Sec.~\ref{sec;2}, we discuss the scales, 
the counting rules, and the  effective Lagrangian for the 
$\Lambda\Lambda d$ and $\Lambda\Lambda\alpha$ systems,
and display the expression of the dressed propagators
for the two-body composite states and the integral
equations for the three-body part. In Sec.~\ref{sec;3}, 
we examine the limit cycle behavior of the integral equations 
in the asymptotic limit for the systems. In Sec.~\ref{sec;4} 
the numerical results and 
in Sec.~\ref{sec;5} a discussion and the conclusions of the work
are presented.

\section{Formalism}
\label{sec;2}

\subsection{Scales, counting rules, and effective Lagrangian}

To construct an EFT for 
$\nuclide[4][\Lambda\Lambda]{H}$ 
as the $\Lambda\Lambda d$ system,
we treat the deuteron field
as a cluster field, i.e., like an elementary field.
Thus the deuteron binding energy, 
$B_2\simeq 2.22$~MeV, is chosen as the high energy scale.
The large (high momentum) scale $\Lambda_H$ of the system
is the deuteron binding momentum,
$\gamma =\sqrt{m_N^{} B_2}\simeq 45.7$~MeV, where $m_N^{}$ is the
nucleon mass. 
We take as the typical momentum ($Q$) 
of the reaction the $\Lambda$ particle separation momentum 
from the hypertriton, which is defined 
by $\gamma_{\Lambda d}^{} = \sqrt{2\mu_{\Lambda d}^{} B_\Lambda} 
\simeq 13.5\pm 2.6$~MeV,
where $\mu_{\Lambda d}^{}$ is the reduced mass of the $\Lambda d$ system
and $B_\Lambda$ is the $\Lambda$ particle separation energy
from the hypertriton, 
$B_\Lambda^{\rm expt.} \simeq 0.13\pm 0.05$~MeV~\cite{JBKK73}.
Then our expansion parameter is
$Q/\Lambda_H\sim \gamma_{\Lambda d}^{}/\gamma \simeq 1/3$, which supports
our expansion scheme.

To construct an EFT for $\nuclide[6][\Lambda\Lambda]{He}$ as 
the $\Lambda\Lambda\alpha$ system,
we treat the $\alpha$ particle field as an elementary field.
The binding energy of the $\alpha$ particle is $B_4 \simeq 28.3$~MeV and its
first excited state has the quantum numbers ($J^\pi = 0^+, I=0$) and the
excitation energy of $E_1 \simeq 20.0$~MeV, which is between the energy gap of
$\nuclide[3][]{H}$-$p$ (19.8~MeV) from the ground state energy and that of
$\nuclide[3][]{He}$-$n$ (20.6~MeV).
Thus the large momentum scale of the $\alpha$-cluster theory is
$\Lambda_H \simeq \sqrt{2\mu E_1}\sim 170$~MeV where $\mu$ is the reduced
mass of the ($\nuclide[3][]{H},p)$ system or the $(\nuclide[3][]{He},n)$ system
so that $\mu \simeq \frac34 m_N^{}$. 
One can see that the mixing of the $\Xi N$ channel in the $\Lambda\Lambda$
interaction becomes an irrelevant degree of freedom 
because of the mass difference $\sim 23$~MeV.
On the other hand, we take the binding momentum of $\nuclide[5][\Lambda]{He}$
as the typical momentum scale $Q$ of the theory.
The $\Lambda$ separation energy of $\nuclide[5][\Lambda]{He}$ is
$B_\Lambda \simeq 3.12$~MeV and thus the binding momentum of
$\nuclide[5][\Lambda]{He}$ as the $\Lambda\alpha$ cluster system is
$\gamma_{\Lambda\alpha}^{} = \sqrt{2\mu_{\Lambda\alpha}^{} B_\Lambda}$,
where $\mu_{\Lambda\alpha}^{}$ is the reduced mass of the $\Lambda\alpha$
system.
This leads to $\gamma_{\Lambda\alpha}^{} \simeq 73.2$~MeV and
thus our expansion parameter is
$Q/\Lambda_H \sim \gamma_{\Lambda\alpha}^{}/\Lambda_H \simeq 0.4$.

We formally employ counting rules 
suggested by Bedaque and van Kolck for two-body systems
in pionless EFT~\cite{bv-plb98,v-lnp98}.\footnote{
Counting rules known as 
the KSW counting rules suggested by Kaplan, Savage, and Wise~\cite{KSW} 
are for two nucleons and pions in chiral EFT. 
} 
At LO the $\Lambda\Lambda$, $\Lambda d$, or $\Lambda\alpha$ 
bubble diagram is resummed up to infinite order. 
The parameters in the renormalized composite propagators
are determined by the scattering length and the binding
momenta corresponding to the two-body binding energies.
For the three-body systems, we employ the counting rules suggested
by Bedaque, Hammer, and van Kolck~\cite{BHvK}. When the system
exhibits the limit cycle, the three-body contact interaction
is promoted to LO in order to renormalize the cyclic singularity 
emerging in the system.

An effective Lagrangian consists of relevant degrees of freedom
at low energies, for which the most general terms are constructed 
based on symmetries of the mother theory, and is perturbatively 
expanded in terms of the number of derivatives~\cite{weinberg}.
Because the typical momentum scales of the systems we consider
are much smaller than the pion mass, the pions fields 
are considered to be heavy degrees of freedom~\cite{aetal-plb04}.
In addition, instead of the chiral limit, 
the unitary limit can be chosen
to describe the three-body
systems~\cite{ah-prc12,ando-fbs13}, 
in which bound states of the three-body systems
can appear as the Efimov states.
While we do not take the unitary limit as LO 
in the present work, and thus all scattering lengths are finite.
Explicit expressions of the effective Lagrangian for the 
$\Lambda\Lambda d$ and $\Lambda\Lambda\alpha$ systems at LO
are presented in Refs. \refcite{dLamH4} and \refcite{dLamHe6}, 
respectively.

\subsection{Two-body part}

Two two-body parts exist in each of the 
$\Lambda\Lambda d$ and $\Lambda\Lambda\alpha$ three-body systems
where we consider only $S$-waves for the two-body interactions at LO. 
Because the $\Lambda\Lambda$ part is common in both systems, 
three two-body parts, 
$\Lambda\Lambda$ in $^1S_0$ channel, 
$\Lambda d$ in ${}_\Lambda^3$H channel, and 
$\Lambda \alpha$ in ${}_\Lambda^5$He channel, are relevant.
The renormalized dressed composite propagators for the 
two-body parts are presented below. 

\begin{figure*}[t]
\begin{center}
\includegraphics[width=0.80\textwidth]{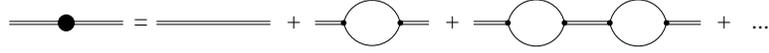}
\caption{
Diagrams for dressed dibaryon propagator.
In the right hand side, the double solid line represents 
the bare dibaryon propagator and
the single solid line denotes the $\Lambda$ propagator.
}
\label{fig;dressed-propagator}
\end{center}
\end{figure*}

The diagrams for the dressed $\Lambda\Lambda$ dibaryon propagator
are depicted in Fig.~\ref{fig;dressed-propagator}.
The dressed $\Lambda\Lambda$ dibaryon propagator at LO 
is obtained as~\cite{dLamH4,BHvK}\footnote{
In Refs.~\refcite{bs-npa01,ah-prc05} the effective range correction
is resummed and included in the propagator.
}
\begin{equation}
D_s(p_0,\vec{p})  
= \frac{4\pi}{m_\Lambda y_s^2}
\frac{1}{
\frac{1}{a_{\Lambda\Lambda}}
- \sqrt{ -m_\Lambda p_0 + \frac14\vec{p}^2
-i\epsilon
}
-i\epsilon
}\,,
\end{equation}
where $p_0$ and $\vec{p}$ are off-shell energy and three momentum,
$m_\Lambda$ is the $\Lambda$ hyperon mass, $a_{\Lambda\Lambda}$
is the $S$-wave $\Lambda\Lambda$ scattering length, 
and $y_s$ is the coupling constant of dibaryon to two $\Lambda$ fields, 
$y_s = -\frac{2}{m_\Lambda}\sqrt{\frac{2\pi}{r_{\Lambda\Lambda}}}$;
$r_{\Lambda\Lambda}$ is the effective range of the $S$-wave $\Lambda\Lambda$
scattering.
We note that it is not necessary to have the effective range, 
$r_{\Lambda \Lambda}$,
at LO. The $r_{\Lambda \Lambda}$ dependence in 
$y_s$ disappears in on-shell quantities, such as a scattering amplitude,
while it remains in intermediate quantities, such as 
$D_s(p_0,\vec{p})$.

\begin{figure*}[t]
\begin{center}
\includegraphics[width=0.80\textwidth]{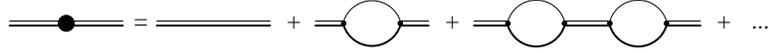}
\caption{
Diagrams for dressed hypertriton propagator as the $\Lambda d$ system.
In the right-hand side, the solid line denotes the $\Lambda$ hyperon while the
thick solid line represents the deuteron.
The bare $t$ field as a $\Lambda d$ composite state in hypertriton channel is denoted by
the double (thin and thick) solid line.
}
\label{fig;dressed-t-propagator}
\end{center}
\end{figure*}

In Fig.~\ref{fig;dressed-t-propagator}, diagrams for 
the dressed hypertriton propagator as the $\Lambda d$ system are depicted.   
The renormalized dressed hypertriton propagator is given as~\cite{dLamH4} 
\begin{equation}
D_t(p_0,\vec{p}) 
= \frac{2\pi}{\mu_{\Lambda d}y_t^2}
\frac{1}{
\gamma_{\Lambda d} - \sqrt{
-2\mu_{\Lambda d}\left(
p_0 
- \frac{1}{2(m_d + m_\Lambda)}\vec{p}^2 
\right)
-i\epsilon
}
-i\epsilon
}\,,
\label{eq;hypertriton_propagator}
\end{equation}
where $\mu_{\Lambda d}$ is the reduced mass of the $\Lambda d$ system,
$m_d$ is the deuteron mass, $\gamma_{\Lambda d}$ is the binding
momentum of the hypertriton.
In addition, $y_t$ is the coupling constant of bare composite hypertriton
filed to the $\Lambda$ hyperon and the deuteron fields, 
$y_t = - \frac{1}{\mu_{\Lambda d}}\sqrt{\frac{2\pi}{r_{\Lambda d}}}$;
$r_{\Lambda d}$ is the effective range of the $S$-wave $\Lambda d$ 
scattering.\footnote{
See footnote \ref{footnote;r}.
}

The expression of the dressed composite ${}_\Lambda^5$He propagator 
of the $\Lambda\alpha$ system is 
similarly represented as that in Eq.~(\ref{eq;hypertriton_propagator})
in terms of the reduced mass $\mu_{\Lambda\alpha}$,
the $\alpha$ particle mass $m_\alpha$, 
and the binding momentum $\gamma_{\Lambda\alpha}$ of ${}_\Lambda^5$He 
as the $\Lambda\alpha$ system. 
One can find it in Eq.~(8) in Ref.~\refcite{dLamHe6}.

\subsection{Three-body part}

For the $\Lambda\Lambda d$ system, 
we derive integral equations for the $S$-wave $\Lambda$ hyperon 
hypertriton scattering.
Because the $\Lambda$ hyperon and the hypertriton are both spin-1/2
there are two total spin states, spin-0 and 1, where we 
consider only $S$-waves between two constituents of the three-body 
system at LO.

\begin{figure}[t]
\begin{center}
\includegraphics[width=0.45\columnwidth]{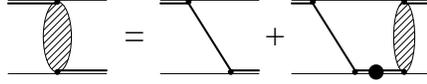}
\caption{
Diagrams of the integral equation for $S$-wave scattering of hypertriton and $\Lambda$
for the spin singlet channel.
See the caption of Fig.~\ref{fig;dressed-t-propagator} as well.
}
\label{fig;singlet-IE}
\end{center}
\end{figure}

In Fig.~\ref{fig;singlet-IE}, diagrams for the $S$-wave $\Lambda$ hyperon 
and hypertriton scattering in the spin-0 channel are depicted.  
We have a single integral equation in terms of the elastic scattering
amplitude $t(p,k;E)$ in momentum space where $p=|\vec{p}|$ 
($k=|\vec{k}|$) is off-shell outgoing (on-shell incoming) relative 
three momentum in C.M. frame and $E$ is the total energy of the system 
as~\cite{dLamH4} 
\begin{eqnarray}
t(p,k;E) &=& -3 K_{(a)}(p,k;E) 
\nonumber \\ &&
+ \frac{1}{2\pi^2}\int^{\Lambda_c}_0 dll^2
3K_{(a)}(p,l;E) D_t\left(
E - \frac{\vec{l}^2}{2m_\Lambda},\vec{l}
\right) t (l,k;E)\,,
\label{eq;t_equation}
\end{eqnarray}
where $K_{(a)}(p,l;E)$ is one-deuteron exchange interaction given as 
\begin{equation}
K_{(a)}(p,l;E) = 
\frac{m_dy_t^2}{6pl}
\ln\left(
\frac{p^2 + l^2 + \frac{2\mu_{\Lambda d}}{m_d} -2\mu_{\Lambda d} E}
     {p^2 + l^2 - \frac{2\mu_{\Lambda d}}{m_d} -2\mu_{\Lambda d} E}
\right)\,,
\end{equation}
and a sharp momentum cutoff $\Lambda_c$ has been introduced 
in the equation.
The total energy $E$ for the scattering is 
$E = - \frac{\gamma_{\Lambda d}^2}{2\mu_{\Lambda d}}
+ \frac{1}{2\mu_{\Lambda(\Lambda d)}} k^2$ where
$\mu_{\Lambda(\Lambda d)} = m_\Lambda(m_\Lambda + m_d)/(2m_\Lambda + m_d)$. 

\begin{figure*}[t]
\begin{center}
\includegraphics[width=0.90\textwidth]{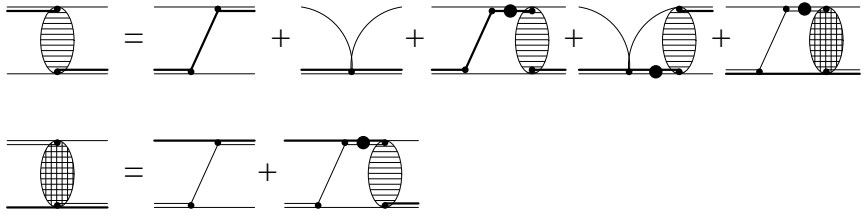}
\caption{
Diagrams of coupled integral equations for
$S$-wave scattering of hypertriton and $\Lambda$
for spin triplet channel.
See the captions of Figs.~\ref{fig;dressed-propagator}
and \ref{fig;dressed-t-propagator} as well.
}
\label{fig;triplet-integral-equations}
\end{center}
\end{figure*}

In Fig.~\ref{fig;triplet-integral-equations}, diagrams for the 
$S$-wave $\Lambda$ hyperon and hypertriton scattering in spin-1 channel
are depicted.  
We have coupled integral equations in terms of two amplitudes
$a(p,k;E)$ and $b(p,k;E)$ where $a(p,k;E)$ is the amplitude 
in elastic channel and $b(p,k;E)$ is that in inelastic 
channel from the $\Lambda$ hyperon and hypertriton cluster 
channel to the $\Lambda\Lambda$ dibaryon and deuteron cluster 
channel as~\cite{dLamH4}~\footnote{
Because the second integral equation in Eq.~(\ref{eq;b_equation})
 depends only on $a(p,k;E)$,
one can use it in the first integral equation in Eq.~(\ref{eq;a_equation})
leading to a single integral equation for $a(p,k;E)$.
} 
\begin{eqnarray}
a(p,k;E) &=& 
K_{(a)}(p,k;E) - \frac{g_1(\Lambda_c)}{\Lambda_c^2}
\nonumber \\ &&
- \frac{1}{2\pi^2}\int^{\Lambda_c}_0 dll^2 \left[
K_{(a)}(p,k;E) - \frac{g_1(\Lambda_c)}{\Lambda_c^2}
\right]
D_t\left( 
E-\frac{\vec{l}^2}{2m_\Lambda},\vec{l}
\right) a(l,k;E)
\nonumber \\ &&
- \frac{1}{2\pi^2}\int^{\Lambda_c}_0 dll^2 
K_{(b1)}(p,k;E) 
D_s\left( 
E-\frac{\vec{l}^2}{2m_\Lambda},\vec{l}
\right) b(l,k;E)\,,
\label{eq;a_equation}
\\
b(p,k;E) &=& K_{(b2)}(p,k;E)
\nonumber \\ &&
- \frac{1}{2\pi^2}\int^{\Lambda_c}_0 dll^2 
K_{(b2)}(p,k;E) 
D_t\left( 
E-\frac{\vec{l}^2}{2m_\Lambda},\vec{l}
\right) a(l,k;E)\,,
\label{eq;b_equation}
\end{eqnarray}
where $g_1(\Lambda_c)$ is the coupling constant of the three-body 
contact interaction in the elastic channel. 
As will be seen below the system in 
this channel exhibits a limit cycle and the strength of the 
coupling $g_1(\Lambda_c)$ is fixed as a function of $\Lambda_c$.
$K_{(b1)}(p,l;E)$ and $K_{(b2)}(p,l;E)$ are one-$\Lambda$ hyperon
exchange interactions given as 
\begin{eqnarray}
K_{(b1)}(p,l;E) &=& -\sqrt{\frac23}\frac{m_\Lambda y_sy_t}{2pl}
\ln\left(
\frac{p^2 + \frac{m_\Lambda}{2\mu_{\Lambda d}}l^2 + pl - m_\Lambda E}
     {p^2 + \frac{m_\Lambda}{2\mu_{\Lambda d}}l^2 - pl - m_\Lambda E}
\right)\,,
\\
K_{(b2)}(p,l;E) &=& -\sqrt{\frac23}\frac{m_\Lambda y_sy_t}{2pl}
\ln\left(
\frac{\frac{m_\Lambda}{2\mu_{\Lambda d}}p^2 + l^2 + pl - m_\Lambda E}
     {\frac{m_\Lambda}{2\mu_{\Lambda d}}p^2 + l^2 - pl - m_\Lambda E}
\right)\,.
\end{eqnarray}

For the $\Lambda\Lambda\alpha$ system to describe the bound state
of ${}_{\Lambda\Lambda}^{\ \ 6}$He, we have similar coupled integral 
equations to those in Eqs.~(\ref{eq;a_equation}) and (\ref{eq;b_equation}).
Explicit expression of the equations can be found in Eq.~(9) in 
Ref.~\refcite{dLamHe6}.

\section{Limit Cycle in the Systems}
\label{sec;3}

In this section, 
we derive equations in the asymptotic limit 
from the integral equations obtained 
in the previous section, 
and make a simple examination whether 
the systems exhibit a limit cycle or not.
In the asymptotic limit,
$l\sim p \gg E, k, 1/a_{\Lambda\Lambda}, \gamma_{\Lambda d} 
(\gamma_{\Lambda\alpha})$, 
the scales in the equations disappear 
and those equations have no scale dependence.
The scale invariance in the asymptotic limit suggests that the 
amplitudes must exhibit a power-law behavior as~\cite{BHvK} 
\begin{equation}
t(p), \ a(p) \sim p^{-1+s}\,,
\end{equation}
where the $k$ and $E$ dependence in the amplitudes is dismissed. 
Then, by performing a Mellin transformation~\cite{jp-fbs13} 
to the homogeneous part of Eq.~(\ref{eq;t_equation}) 
and Eqs.~(\ref{eq;a_equation},\ref{eq;b_equation}) in the limit
we obtain~\cite{dLamHe6}
\begin{eqnarray}
1 &=& C_0 I_1(s) \,,
\label{eq;C0I1}
\\  
1 &=& C_1 I_1(s) + C_2 I_2(s)I_3(s)\,,
\label{eq;C1I1}
\end{eqnarray}
with
\begin{eqnarray}
C_0 &=& - \frac{1}{2\pi}\frac{m_d}{\mu_{\Lambda d}}
\sqrt{\frac{\mu_{\Lambda(\Lambda d)}}{\mu_{\Lambda d}}}\,,
\\ 
C_1 &=& \frac{1}{6\pi}\frac{m_d}{\mu_{\Lambda d}}
\sqrt{
\frac{\mu_{\Lambda(\Lambda d)}}{\mu_{\Lambda d}}
}\,,
\ \ 
C_2 = \frac{\sqrt2}{3\pi^2 }
\frac{
\sqrt{m_\Lambda \mu_{d(\Lambda\Lambda)}\mu_{\Lambda (\Lambda d)}}
}{ \mu_{\Lambda d}^{3/2}} \,,
\end{eqnarray}
where $\mu_{d(\Lambda\Lambda)} = 2m_\Lambda m_d/(2m_\Lambda + m_d)$,
and
\begin{eqnarray}
I_1(s) &=& 
\frac{2\pi}{s} \frac{\sin[s\sin^{-1}\left(\frac12a\right)]}{
\cos\left(\frac{\pi}{2}s\right)},
\label{eq;I1}
\\
I_2(s) 
&=& \frac{2\pi}{s}  \frac{1}{b^{s/2}}
\frac{\sin[s\cot^{-1}\left(\sqrt{4b-1}\right)]}
{\cos\left(\frac{\pi}{2}s\right)} ,
\label{eq;I2}
\\
I_3(s) 
&=& \frac{2\pi}{s} b^{s/2}
\frac{\sin[s\cot^{-1}\left(\sqrt{4b-1}\right)]}
{\cos\left(\frac{\pi}{2}s\right)} ,
\label{eq;I3}
\end{eqnarray}
and $a= \frac{2\mu_{\Lambda d}}{m_d}$ and 
$b= \frac{m_\Lambda}{2\mu_{\Lambda d}}$.
If the 
solutions of the equation in Eq.~(\ref{eq;C0I1}) or (\ref{eq;C1I1}) 
are real, the physical solution of the amplitude should satisfy 
the condition that the half-off-shell amplitude converges 
in the asymptotic limit.
On the other hand, 
if the solutions of the equation are complex, e.g., $s=\pm i s_0$,
then the amplitude exhibits a cyclic behavior 
as $t(p),a(p)\sim e^{\pm i s_0\ln(p)}$  
in the limit where $p\to \infty$.
Thus it is necessary to introduce a counter term
to renormalize the cyclic divergence 
and fix the counter term by choosing a renormalization point.

The solutions of $s$ in Eq.~(\ref{eq;C0I1}) are real and we have 
$s = \pm 2.0$, $\pm 2.0838\cdots$,  $\pm 5.3227\cdots$, $\pm 6.8665\cdots$,
$\cdots$. 
Thus the $\Lambda\Lambda d$
system in the spin-0 channel will not exhibit 
a limit cycle and 
it is not necessary to introduce 
a three-body contact interaction
for renormalization. 
It is expected that the system in this channel is not sensitive to a 
three-body contact interaction. 
The scattering amplitude 
for $S$-wave $\Lambda$ hyperon and hypertriton scattering 
in the spin-0 channel may be well described by the two-body
interaction at LO. 
The scattering length for the process is to be estimated
in the next section.

The solutions of $s$ in Eq.~(\ref{eq;C1I1}) are
imaginary and we have $s=\pm i s_0$ where
\begin{equation}
s_0 = 0.4492\cdots\,.
\label{eq;s0_4}
\end{equation}  
In addition, the multiplicative factor of the discrete scaling symmetry 
becomes $e^{\pi/s_0}\simeq 1.09\times 10^3$.
Thus the $\Lambda\Lambda d$ system in the spin-1 channel will exhibit
a limit cycle and the three-body contact interaction would be promoted to LO. 
However, it may not be easy to observe the cyclic pattern 
in the system because the typical scale is $\gamma_{\Lambda d}\simeq
13.5$~MeV and the next scale in the discrete scaling symmetry appears
at $\gamma_{\Lambda d}^{(2)} =e^{\pi/s_0}\gamma_{\Lambda d} 
\sim 15 \times 10^3$~MeV, which is
much larger than the hard scale of the theory,
$\Lambda_H\sim 50$~MeV.

In the case of the $\Lambda\Lambda\alpha$ system, on the other hand,
we have the same
equation as that in Eq.~(\ref{eq;C1I1}) but different values of the 
coefficients, $C_1$ and $C_2$ and the parameters $a$ and $b$ in the 
functions $I_1(s)$, $I_2(s)$, and $I_3(s)$ in Eqs.~(\ref{eq;I1}),
(\ref{eq;I2}), and (\ref{eq;I3})~\cite{dLamHe6}, which are
\begin{equation}
C_1 = \frac{1}{2\pi}\frac{m_\alpha}{\mu_{\Lambda\alpha}}
\sqrt{\frac{\mu_{\Lambda(\Lambda\alpha)}}{\mu_{\Lambda\alpha}}}\,,
\ \ 
C_2 = \frac{\sqrt{2m_\Lambda \mu_{\Lambda(\Lambda\alpha)}
\mu_{\alpha(\Lambda\Lambda)}}}{\pi^2\mu_{\Lambda\alpha}^{3/2}}\,,
\end{equation} 
where $\mu_{\Lambda(\Lambda\alpha)} 
= m_\Lambda(m_\Lambda+m_\alpha)/(2m_\Lambda + m_\alpha)$ 
and $\mu_{\alpha(\Lambda\Lambda)}=2m_\Lambda m_\alpha/(2m_\Lambda+m_\alpha)$,
and $a= 2\mu_{\Lambda\alpha}/m_\alpha$ and $b=m_\Lambda/(2\mu_{\Lambda\alpha})$.
The solutions of the equation are imaginary as well 
and we have $s=\pm is_0$ where
\begin{equation}
s_0 = 1.0496\cdots \,.
\label{eq;s0_6}
\end{equation}
The multiplicative factor becomes $e^{\pi/s_0}\simeq 19.9$,
which is similar to that in the case of three nucleons in triton channel,
$e^{\pi/s_0}\simeq 22.7$, where $s_0\simeq 1.00624$~\cite{BHV99,AB10b}.
Thus the $\Lambda\Lambda\alpha$ system in the spin-0 channel will
exhibit a limit cycle and 
a three-body contact interaction should be
included at LO. The typical scale of the system is 
$\gamma_{\Lambda\alpha}\simeq 73.2$~MeV and the next scale in the discrete
scaling symmetry will appear at 
$\gamma_{\Lambda\alpha}^{(2)}=e^{\pi/s_0}\gamma_{\Lambda\alpha}
\simeq 1.5\times 10^3$~MeV, which is about 10 times larger than the 
hard scale of the theory, $\Lambda_H\simeq 170$~MeV. 
Thus it would not be easy to see the second one of the discrete
scaling symmetry in the $\Lambda\Lambda\alpha$ system either.
Though it may be difficult to observe the cyclic pattern in the systems,
the first bound state appearing in
the limit cycle may correspond to those of 
$\nuclide[4][\Lambda\Lambda]{H}$ 
and $\nuclide[6][\Lambda\Lambda]{He}$.
That is to be studied in the next section.

\section{Numerical Results}
\label{sec;4}

In this section we review our numerical results presented 
in Refs.~\refcite{dLamH4,dLamHe6}.
We discuss the numerical results 
for the scattering length of the $S$-wave $\Lambda$ hyperon 
and hypertriton in the spin-0 channel, in which 
a limit cycle does not appear, 
and for the bound states of ${}_{\Lambda\Lambda}^{\ \ 4}$H
and ${}_{\Lambda\Lambda}^{\ \ 6}$He as 
the $\Lambda\Lambda d$ system in the spin-1 channel and
the $\Lambda\Lambda\alpha$ system in the spin-0 channel, 
in which the limit cycle appears.

\subsection{Scattering channel not exhibiting limit cycle}

The integral equation at LO in Eq.~(\ref{eq;t_equation}) can be
fixed by the masses and the two effective range parameters,
$\gamma_{\Lambda d}$ and $r_{\Lambda d}$, where
$\gamma_{\Lambda d}= 13.5$~MeV and 
$r_{\Lambda d}=2.3\pm 0.3$~fm~\cite{Hammer01}.
\footnote{
As mentioned above, 
it is not necessary to have the effective range, $r_{\Lambda d}$,
at LO. In our calculation, the $r_{\Lambda d}$ dependence indeed
disappears in the scattering amplitude, $T(k,k)$, while it remains 
in the intermediate quantities, such as $K_{(a)}(p,k;E)$, 
$D_t(p_0,\vec{p})$, and $Z_{\Lambda d}$. 
\label{footnote;r}
}
The scattering length $a_0$ of the $S$-wave $\Lambda$ hyperon and hypertriton
scattering
in the spin-0 channel is calculated by using the formulae,
\begin{equation}
a_0 = - \frac{\mu_{\Lambda(\Lambda d)}}{2\pi} T(0,0)\,.
\end{equation}
In the expression above, $T(k,k)$ is the on-shell scattering amplitude 
given by
\begin{equation}
T(k,k) = \sqrt{Z_{\Lambda d}}\,t(k,k;E)\sqrt{Z_{\Lambda d}}\,,
\end{equation}
where $Z_{\Lambda d}$ is the wavefunction normalization factor of the 
hypertriton as the $\Lambda d$ system, 
$Z_{\Lambda d}= \gamma_{\Lambda d}r_{\Lambda d}$. 
Thus we obtain~\cite{dLamH4}
\begin{equation}
a_0 = 16.0\pm 3.0\,\ \mbox{\rm fm}\,.
\end{equation}
No experimental datum for the quantity is available 
because there are no enough hypertriton targets for experiment.
The phase shift up to the hypertriton breakup threshold 
is also calculated and displayed in Fig.~6 in Ref.~\refcite{dLamH4}.

\subsection{Binding channels exhibiting limit cycle }

\subsubsection{${}_{\Lambda\Lambda}^{\ \ 4}$H as $\Lambda\Lambda d$ system }

The $\Lambda\Lambda d$ system in the spin-1 channel 
described by the coupled integral equations at LO 
can be determined by three constants, 
$\gamma_{\Lambda d}$, $a_{\Lambda\Lambda}$,
and $g_1(\Lambda_c)$. As mentioned, 
$\gamma_{\Lambda d}=13.5$~MeV whereas 
the value of $a_{\Lambda\Lambda}^{}$ 
is experimentally deduced from the
${}^{12}$C($K^-,K^+\Lambda\Lambda X)$
reaction~\cite{YAAF07}, which leads to
$a_{\Lambda\Lambda}^{} = -1.2 \pm 0.6$~fm~\cite{GHH11}.
In addition,
the data for the Au+Au collisions at the Relativistic Heavy Ion Collider~\cite{STAR13}
are analyzed and $a_{\Lambda\Lambda}^{} \ge -1.25$~fm 
is reported
in Ref.~\refcite{ExHIC-13}.
To fix the coefficient of the three-body contact interaction
$g_1(\Lambda_c)$ one needs to have some three-body datum, 
but there are
no available experimental data at the present moment.
We make use of the results from the model calculations 
in which the formation of the ${}_{\Lambda\Lambda}^{\ \ 4}$H bound
state is reported~\cite{FG02,NAM02}. 
We employ three set of the parameters 
of $B_{\Lambda\Lambda}$ and $a_{\Lambda\Lambda}$:
\begin{eqnarray}
\mbox{\rm (I)} && \  B_{\Lambda\Lambda}\simeq 0.2~\mbox{\rm MeV} 
\ \mbox{and} \ a_{\Lambda\Lambda}\simeq -0.5~\mbox{\rm fm}\,,  
\\
\mbox{\rm (II)} && \  B_{\Lambda\Lambda}\simeq 0.6~\mbox{\rm MeV} 
\ \mbox{and} \ a_{\Lambda\Lambda}\simeq -1.5~\mbox{\rm fm}\,,  
\\
\mbox{\rm (III)} && \  B_{\Lambda\Lambda}\simeq 1.0~\mbox{\rm MeV} 
\ \mbox{and} \ a_{\Lambda\Lambda}\simeq -2.5~\mbox{\rm fm}\,.  
\end{eqnarray}

\begin{figure}[t]
\begin{center}
\includegraphics[width=0.6\columnwidth]{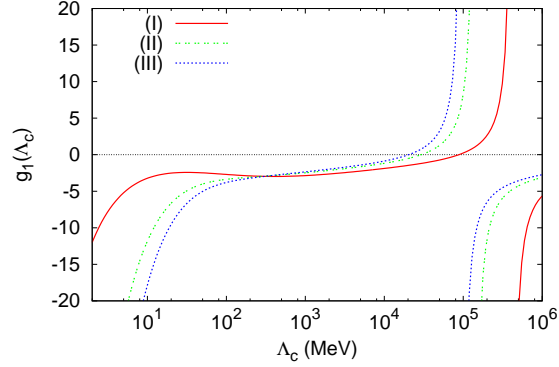}
\caption{
Coupling $g_1^{}(\Lambda_c)$ of three-body contact interaction 
as functions of the cutoff $\Lambda_c$ which produces a bound 
state of $_{\Lambda\Lambda}^{\ \ 4}$H  
with three different sets of $B_{\Lambda\Lambda}$ and $a_{\Lambda\Lambda}^{}$.
See the text for the parameter sets (I), (II), and (III).
}
\label{fig;g1vsLam}
\end{center}
\end{figure}

In Fig.~\ref{fig;g1vsLam}, we plot the curves of $g_1(\Lambda_c)$ as 
functions of the cutoff $\Lambda_c$ where $g_1(\Lambda_c)$ is fixed
by using the three sets of the parameters labeled by (I), (II), and (III)
introduced above.
One can see that the curves of $g_1^{} (\Lambda_c)$ are rather 
mildly varying at 
$\Lambda_c = 10 \sim 10^4$~MeV, and each curve has a singularity at
$\Lambda_c \sim 10^5$~MeV indicating the possibility of the first cycle 
of the limit-cycle. 
This implies that the one-deuteron-exchange and one-$\Lambda$-exchange
interactions for the spin-1 channel contains an attractive (singular) 
interaction at very high momentum, say, $\Lambda_c \sim 10^5$~MeV.
At such a very high momentum, however, 
the applicability of the present theory, a very low
energy EFT, cannot be guaranteed and thus the mechanisms of the formation of
a bound state must have different origins.
We note, on the other hand, that, if we choose $g_1^{} (\Lambda_c) \simeq -2$
or smaller at $\Lambda_c \sim 50$~MeV in the coupled integral equations,
a bound state can be created.
Such a value of $g_1^{} (\Lambda_c)$ is of natural size and may be generated
from the mechanisms of high energy such as $\sigma$-meson exchange or
two-pion exchange near the intermediate range of nuclear force, i.e.,
$\Lambda_c =300 \sim 600$~MeV.

\begin{figure}[t]
\begin{center}
\includegraphics[width=0.6\columnwidth]{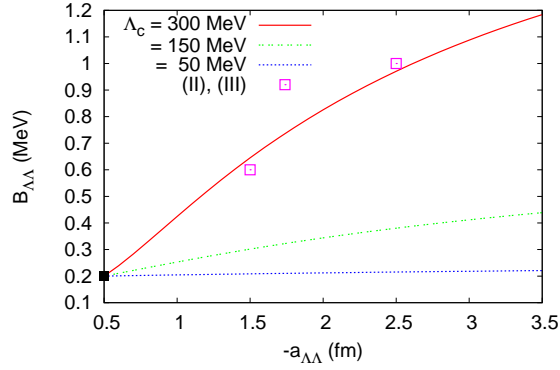}
\caption{
Calculated two-$\Lambda$ separation energy $B_{\Lambda\Lambda}$ from
${}_{\Lambda\Lambda}^{\ \ 4}$H bound state as functions of the scattering 
length $a_{\Lambda\Lambda}^{}$ of the $S$-wave $\Lambda\Lambda$ scattering 
for the ${}^1S_0$ channel with the cutoff values $\Lambda_c=50$, 150, 300~MeV.
The value of $g_1^{} (\Lambda_c)$ of all three curves is fitted
at the point (I):
$B_{\Lambda\Lambda}=0.2$~MeV and $a_{\Lambda\Lambda}^{} = -0.5$~fm,
marked by a filled square.
The points (II) and (III) are also included as blank squares in the figure.
}
\label{fig;BLLvsaLL}
\end{center}
\end{figure}

In Fig.~\ref{fig;BLLvsaLL}, we plot the curves of $B_{\Lambda\Lambda}$
as functions of $a_{\Lambda\Lambda}$.
Here, the coupling $g_1^{} (\Lambda_c)$ is fixed by 
using the parameter set (I), i.e.,
$B_{\Lambda\Lambda}=0.2$~MeV and $a_{\Lambda\Lambda}=-0.5$~fm, which is
marked by a filled square in Fig.~\ref{fig;BLLvsaLL}.
This is achieved with $g_1^{} (\Lambda_c) \simeq -2.48$, $-2.83$, $-2.96$
for $\Lambda_c=50$, $150$, $300$~MeV, respectively.
Once the starting values are fixed, we vary the value of 
$a_{\Lambda\Lambda}^{}$ 
for a fixed value of $\Lambda_c$, which changes the values 
of $B_{\Lambda\Lambda}$.
We then find that the behaviors of the $B_{\Lambda\Lambda}$ curves
as functions of $a_{\Lambda\Lambda}$ are quite sensitive to the values of
the cutoff $\Lambda_c$.
For example, when we choose $\Lambda_c \simeq \Lambda_H$,
i.e., $\Lambda_c=50$~MeV, $B_{\Lambda\Lambda}$ is insensitive to the value of
$a_{\Lambda\Lambda}^{}$ and makes a nearly flat curve as shown 
by the dotted line in Fig.~\ref{fig;BLLvsaLL}.
However, with a larger cutoff value, $\Lambda_c = 300$~MeV, $B_{\Lambda\Lambda}$
strongly depends on $a_{\Lambda\Lambda}^{}$ and we can fairly well
reproduce the $a_{\Lambda\Lambda}^{}$-dependence of $B_{\Lambda\Lambda}$
obtained in the potential model calculation
by Filikhin and Gal~\cite{FG02} or Nemura {\it et al.}~\cite{NAM02}.

\subsubsection{${}_{\Lambda\Lambda}^{\ \ 6}$He 
as $\Lambda\Lambda\alpha$ system}

The $\Lambda\Lambda\alpha$ system in the spin-0 channel described by
the coupled integral equations at LO can also be determined by three 
coupling constants, $\gamma_{\Lambda\alpha}$, $a_{\Lambda\Lambda}$,
and $g(\Lambda_c)$, 
where $g(\Lambda_c)$ is a coupling constant of a three-body contact 
interaction for the $\Lambda\Lambda\alpha$ system~\cite{dLamHe6}. 
As mentioned above, 
the $\Lambda$ separation momentum of $^5_{\Lambda}$He is given by 
$\gamma_{\Lambda\alpha} \simeq 73.2$~MeV, 
there is the 
uncertainty in $a_{\Lambda\Lambda}$,
and  $g(\Lambda_c)$ can be 
fixed by using the experimental data of $B_{\Lambda\Lambda}$ from 
${}_{\Lambda\Lambda}^{\ \ 6}$He, $B_{\Lambda\Lambda}= 6.93\pm 0.16$~MeV.

\begin{figure}[t]
\begin{center}
\includegraphics[width=0.6\columnwidth]{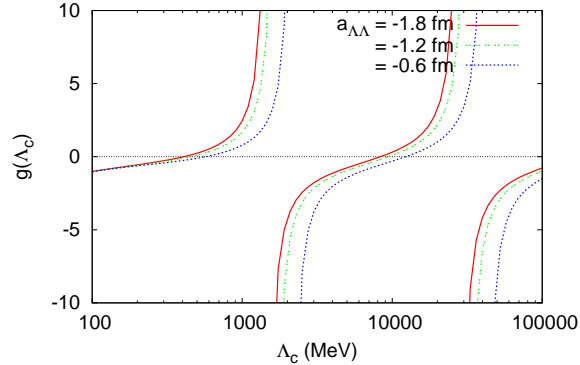}
\caption{
The coupling $g(\Lambda_c)$ as functions of $\Lambda_c$
for $a_{\Lambda\Lambda}^{}=-0.6$, $-1.2$, $-1.8$~fm
where the values of $g(\Lambda_c)$ are fitted by
$B_{\Lambda\Lambda}=6.93$~MeV of $\nuclide[6][\Lambda\Lambda]{He}$.
}
\label{fig;gvsLam}
\end{center}
\end{figure}

In Fig.~\ref{fig;gvsLam}, we plot the curves of $g(\Lambda_c)$
as functions of the cutoff $\Lambda_c$ with three different values
of $a_{\Lambda\Lambda}= - 1.8, -1.2, -0.6$~fm 
so as to reproduce the experimental data
of $B_{\Lambda\Lambda}$. 
The curves are numerically obtained from the homogeneous part of the coupled
integral equations in Eq.~(9) in Ref.~\refcite{dLamHe6}. 
One can see that the curves clearly exhibit a limit cycle and the first
divergence appears at $\Lambda_c \sim 1$~GeV.
In addition, a larger value of $|a_{\Lambda\Lambda}^{}|$ behaves as giving
a larger attractive force and shifts the curves of $g(\Lambda_c)$ 
to the left in Fig.~\ref{fig;gvsLam}.

The value of $s_0^{}$ 
obtained in Eq.~(\ref{eq;s0_6}) can be estimated from the curves 
of the limit cycle of
$g(\Lambda_c)$ in Fig.~\ref{fig;gvsLam}.
The $(n+1)$-th values of $\Lambda_n$ at which $g(\Lambda_c)$ vanishes
can be parameterized as $\Lambda_n = \Lambda_0\exp(n\pi/s_0^{})$
in the discrete scaling symmetry.
By using the second and third vales of $\Lambda_n$ for the three
values of $a_{\Lambda\Lambda}^{}$, we have
$s_0^{} = \pi/\ln(\Lambda_2/\Lambda_1)\simeq 1.05$, which is in a very good
agreement with the value of Eq.~(\ref{eq;s0_6}).
Furthermore, the value of $s_0$ may be checked by using Fig.~52 
in Ref.~\refcite{BH04}
which is a plot of $\exp{(\pi/s_0)}$ versus $m_1/m_3$ for the mass-imbalanced
system where $m_1=m_2\neq m_3$. In our case, $m_1/m_3 = m_\Lambda/m_\alpha
\simeq 0.3$, one finds $\exp(\pi/s_0)\simeq 20$ from the figure
and thus $s_0 \simeq 1.05$. This is another very good agreement with what we
find in Eq.~(\ref{eq;s0_6}).

\begin{figure}[t]
\begin{center}
\includegraphics[width=0.6\columnwidth]{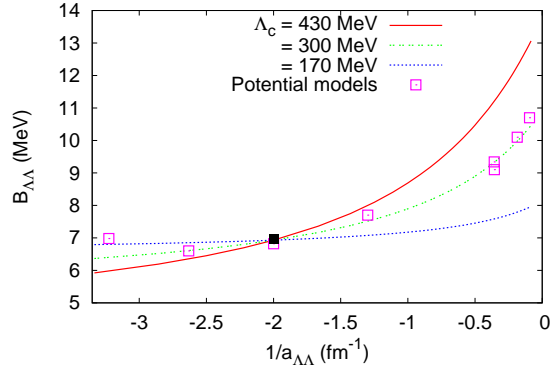}
\caption{
The two-$\Lambda$ separation energy $B_{\Lambda\Lambda}$ as functions
of $1/a_{\Lambda\Lambda}^{}$ for $\Lambda_c=170$, $300$, $430$~MeV, where
$g(\Lambda_c)$ is renormalized at the point of $B_{\Lambda\Lambda}=6.93$~MeV
and $1/a_{\Lambda\Lambda}^{}=-2.0$~fm$^{-1}$ that is marked by a filled square.
Open squares are the results from the potential models in Table 5 of 
Ref.~\cite{FG02b}.
}
\label{fig;BLL6vsaLL}
\end{center}
\end{figure}

In Fig.~\ref{fig;BLL6vsaLL}, we plot the curves of $B_{\Lambda\Lambda}$
as functions of $a_{\Lambda\Lambda}$ where $g(\Lambda_c)$ is renormalized
at the point (marked by the filled square in the figure) of 
$B_{\Lambda\Lambda}=6.93$~MeV and $1/a_{\Lambda\Lambda}= -2.0$~fm$^{-1}$
with three deference values of the cutoff $\Lambda_c = 430$, 300, 170~MeV.
This leads to $g(\Lambda_c) \simeq -0.715$, $-0.447$, $-0.254$ for
$\Lambda_c=170$, $300$, $430$~MeV, respectively.
Open squares are the estimated values from the potential models given
in Table 5 of Ref.~\refcite{FG02b}.
We find that the curves are sensitive to the cutoff values.
When we choose $\Lambda_c = \Lambda_H\simeq 170$~MeV,
the curve for $B_{\Lambda\Lambda}$ is found to be rather insensitive 
to $1/a_{\Lambda\Lambda}$. While the results from
the potential models are remarkably well reproduced by the curve with
$\Lambda_c=300$~MeV.

\section{Discussion and conclusions}
\label{sec;5}

In this article we reviewed the results of our recent works on the 
light double $\Lambda$ hypernuclei, 
${}_{\Lambda\Lambda}^{\ \ 4}$H and
${}_{\Lambda\Lambda}^{\ \ 6}$He, as
the $\Lambda\Lambda d$ and $\Lambda\Lambda\alpha$ three-body systems
investigated in halo/cluster EFT at LO~\cite{dLamH4,dLamHe6}.
We obtained the equations from the integral 
equations in the asymptotic limit to make a simple examination
whether the systems exhibit a limit cycle or not. 
When a limit cycle does not appear in the three-body system, 
it is not necessary to introduce
a three-body contact interaction 
and the system can be described by the two-body interaction. 
Thus we estimate the scattering length $a_0$ for the $S$-wave 
$\Lambda$ hyperon and hypertriton scattering in the spin-0 channel
without an unknown parameter. 
Our theoretical prediction is $a_0=16.0\pm 3.0$~fm.
Though our result cannot be tested due to the lack of any experimental datum,  
the similar case,
the $S$-wave $nd$ scattering in the spin-quartet channel,
may give a positive indication that the theory has a prediction 
power in the channel without exhibiting the limit cycle.
In the three-nucleon system a limit cycle does not appear either 
and the scattering length and the phase shift below the deuteron 
breakup threshold are well described without a three-body
contact interaction~\cite{bv-plb98,Ando13}.
The scattering length $a_4$
is obtained in pionless EFT as $a_4=6.33$~fm~\cite{bv-plb98} which
is in good agreement with the experimental value of
$a_4^{exp.}=6.35\pm 0.02$~fm~\cite{a4exp}.

When the limit cycle appears in the system, 
it is necessary to introduce
the three-body contact interaction 
for renormalization, and 
the bound states appearing in the system 
may sometimes be regarded as an Efimov state.
For the $\Lambda\Lambda d$ system in the spin-1 channel, 
as seen in Fig.~\ref{fig;g1vsLam}, the bound 
state of ${}_{\Lambda\Lambda}^{\ \ 4}$H 
can be formed due to the three-body contact interaction, which stems from
the physics at the high energies, in the range of the cutoff value 
up to $\Lambda_c\sim 10^4$~MeV. The bound state can be formed without
the three-body contact interaction when the cutoff value becomes larger
than $\Lambda_c\sim 2\times 10^4$~MeV.
The value is extremely large compared to 
the hard scale of the $\Lambda\Lambda d$ system, $\Lambda_H\simeq 50$~MeV.
Thus it is hard to regard that of ${}_{\Lambda\Lambda}^{\ \ 4}$H 
as an Efimov state.

On the other hand, the bound state of the $\Lambda\Lambda\alpha$ system,
${}_{\Lambda\Lambda}^{\ \ 6}$He, can be formed 
by choosing the value of the cutoff as $\Lambda_c=450\sim 600$~MeV 
without introducing the three-body contact interaction
where the range of the cutoff dependence comes out of 
the uncertainty of $a_{\Lambda\Lambda}$. 
That is comparable to the 
hard scale of the system, $\Lambda_H\simeq 170$~MeV. 
Thus the bound state for ${}_{\Lambda\Lambda}^{\ \ 6}$He formed
at the first cycle of the limit cycle 
would be regarded as an Efimov state.
Another example for the formation of 
an Efimov state in the three-body system is the triton. 
In pionless EFT calculations, the bound state of the three-nucleon system
is formed without introducing the three-body contact interaction when
one chooses $\Lambda_c\simeq 380$~MeV~\cite{AB10b}. 
That is also comparable to the 
hard scale of the pionless EFT, $\Lambda_H = m_\pi\simeq 140$~MeV.

As discussed in the introduction,
the systems can be described by a small number of the parameters 
in the very low energy EFTs.
There is only one parameter for the $\Lambda\Lambda d$ system 
in the spin-0 channel, whereas there are three 
for the $\Lambda\Lambda d$ system
in the spin-1 channel and the $\Lambda\Lambda \alpha$ system in the 
spin-0 system. In addition, each of the three-body systems is
characterized by the typical and hard scales 
and the presence or absence of the limit cycle.
This is a simple analysis but may provide us nontrivial information
and knowledge about the systems.
It may be interesting to apply the analysis carried out in the present
article to the study of an exotic bound state arising from other three-body
systems~\footnote{
Recently, we applied the analysis to the study of 
a bound state formation in the $nn\Lambda$ system~\cite{aro-15}. 
}.

It may be worth discussing the role of the cutoff value
in the systems.\footnote{
A value of the cutoff $\Lambda_c$ can be arbitrary when renormalizing 
an on-shell physical quantity of the three-body system.
(While we may take the value of $\Lambda_c$ to be smaller than a value
at which an unphysical deeply bound state is formed.) 
In this work, to assure the consistency between the effective degrees of
freedom and the off-shell probe of the loop momentum, 
we assume $\Lambda_c\sim \Lambda_H$. 
}
As mentioned, the correlations between
$B_{\Lambda\Lambda}$ and $a_{\Lambda\Lambda}$ obtained from the 
potential models are well reproduced 
in both the ${}_{\Lambda\Lambda}^{\ \ 4}$H
and ${}_{\Lambda\Lambda}^{\ \ 6}$He states 
when we choose $\Lambda_c = 300$~MeV.
It may indicate that our simple theory 
can probe the scale of the $\Lambda\Lambda$ interactions of 
the potential model calculations.
The value of $\Lambda_c$ is consistent to the scale 
for the long range part of the 
$\Lambda\Lambda$ interaction which consists of the two-pion-exchange,
$2m_\pi\sim 300$~MeV. 
With such a momentum cutoff, however, the two nucleon structure 
of the deuteron and the existence of the excited state of 
the $\alpha$ particle can be probed in the loops.
Because our EFTs at very low energies 
do not have such detailed structures, 
the short range structural mechanism is missing
in that case. 
Such an inconsistency
is in fact common in the cluster model calculations.
 
When we strictly choose 
the cutoff $\Lambda_c$ to be the hard scale
of the systems, $\Lambda_c=\Lambda_H\simeq 50$~MeV for the $\Lambda\Lambda d$
system and $\Lambda_c=\Lambda_H\simeq 170$~MeV for the $\Lambda\Lambda\alpha$
system, on the other hand, 
the curves for $B_{\Lambda\Lambda}$ are less sensitive
to $a_{\Lambda\Lambda}$, compared to the potential models, 
in Figs.~\ref{fig;BLLvsaLL} and 
\ref{fig;BLL6vsaLL}, respectively. 
The contribution from the $\Lambda\Lambda$ two-body part with the 
different cutoff value is eventually compensated by the three-body contact 
interaction after the renormalization of the three-body quantity
in each of the systems. 
In addition, the scale of $1/a_{\Lambda\Lambda}$ is 
$|1/a_{\Lambda\Lambda}| = 330\sim 110$~MeV, which 
is comparable to the hard scales of the systems.
Thus some portion of the $\Lambda\Lambda$ interaction is carried
by the three-body contact term in the very low energy EFTs 
and is not easily disentangled from it.

\section*{Acknowledgements}

The author would like to thank Y.~Oh, G.-S. Yang, and U. Raha
for collaborations. This work was 
supported by the Basic Science Research Program through the National
Research Foundation of Korea funded by the Ministry of Education 
of Korea under Grant No.\ NRF-2012R1A1A2009430. 
This work was also supported in part by the Ministry of Science, 
ICT, and Future Planning (MSIP) and the National Research Foundation 
of Korea under Grant No.\ NRF-2013K1A3A7A06056592
(Center for Korean J-PARC Users).

\end{document}